\documentclass[12pt]{iopart}
\usepackage{amsfonts, amssymb, amsthm}

\newcommand{\sect}[1]{\setcounter{equation}{0}\section{#1}}
\newcommand{\subsect}[1]{\subsection{#1}}

\def\be{\begin{equation}}
\def\ee{\end{equation}}
\def\ba{\begin{array}}
\def\ea{\end{array}}
\def\bea{\begin{eqnarray}}
\def\eea{\end{eqnarray}}
\def\bea*{\begin{eqnarray*}}
\def\eea*{\end{eqnarray*}}
\def\hbar{\mathchar '26\mkern -9muh}     
\def\k#1{#1^{(k)}}

\def\ket#1{|#1 \rangle}

\def\1#1{\left( \frac\omega{\pi\hbar} \right)^{#1}}

\def\PE#1{[#1]}       						

\def\tr{{\rm tr\,}}                  
\def\h{{\cal H}}             
\def\Om{\Omega}              
\def\om{\omega}                     
\def\p{\psi}                        
\def\f{\varphi}                     
\def\th{\theta}                     

\def\R{{\mathbb R}}

\def\Z{{\mathbb Z}}                    

\begin{document}
\
\title[Quantum Mechanics on the cylinder]
{Quantum Mechanics on the cylinder}

{\author{Jos\'e A. Gonz\'alez\footnote{E-mail:  jagonzal$@$fta.uva.es},
Mariano A del Olmo\footnote{E-mail: olmo@fta.uva.es} and 
Jaromir Tosiek\footnote{E-mail: tosiek@ck-sg.p.lodz.pl}
\footnote{{On leave of absent
from Technical University of  Lodz (Poland).}}}}
\address{Departamento de F\'{\i}sica Te\'{o}rica, Universidad de Valladolid\\
47011 Valladolid, Spain}

\begin{abstract} 
\noindent
A new approach   to deformation  quantization on the
cylinder considered as phase space is presented. The method is based on the standard Moyal
formalism for
$\R^2$ adapted  to $S^1\times \R$ by the Weil--Brezin--Zak transformation.  The results
are compared with other solutions of this problem presented by   Kasperkovitz and
Peev [{\em Ann. Phys.} {\bf 230}, 21  (1994)]  and by Pleba\'nski and collaborators [{\em
Acta Phys. Pol.} {\bf B 31}, 561 (2000)]. The equivalence of these three methods is proved.
\end{abstract}

To be published in {\JOB} (2003).
\maketitle

\sect{Introduction}

Mathematical formalisms of classical and quantum physics are completely
different. In the first one physical systems are
described on differential manifolds, in the second one linear operators and vectors on
Hilbert spaces are the suitable objects to describe quantum systems.  However, there are formal
analogies among them, especially between quantum mechanics and classical statistical
mechanics.  Thus, since the beginning of  quantum mechanics  the question appeared if 
states  and observables could be expressed in terms of functions on some phase space, as it
happens in classical theory.  Such a
formalism  would be more intuitive and would enlighten the relation between
classical and quantum mechanics.

Moyal~\cite{MO49} basing on
previous works by Weyl~\cite{WY31}, Wigner~\cite{WI32} and Groenewold~\cite{GW46}
presented quantum mechanics as a statistical theory (see also \cite{GR74,AH81,gad}).
Unfortunately, his method can be applied only to systems with Euclidean configuration
space. 
Quantum states are represented by normalized functions on a phase space (Wigner
functions)  but these functions cannot be interpreted as probability density since they 
are not always positive.  The domains of Wigner functions are low bounded on the contrary to
the classical theories (for instance, the distribution $\delta (q-q') \;\delta (p-p') $
that corresponds to a classical   particle in $\R^2$ cannot be a Wigner
function). These facts  reflect the Heisenberg uncertainty principle. 

With respect to 
observables, they are represented by real functions on phase space. However, their classical
commutative product has to be substituted by a noncommutative multiplication  
($*$--product). The  Moyal bracket (the analog of the commutator of operators), which
characterizes the Lie algebra of observables,  is different, in general, from   the
classical Poisson bracket.

On the other hand, quantum mechanics formalism on phase space can be defined in an
independent way, without any explicit mention  to the Weyl correspondence, like an
algebra  of functions defined on phase space with the Moyal product as fundamental
element. This was the approach followed by Berezin~\cite{BE74,BE75} in his quantization
method. The development of this idea carried to a new theory: deformation quantization
\cite{BF78}, where the quantum formalism is obtained by deforming the
algebra of classical observables by the introduction of a new noncommutative product
($*$--product) between the functions on phase space.  In some sense, this theory
represents  respect to Moyal's formalism an analog role  to the geometric quantization
(see for instance \cite{SO70,WO80} and references therein) respect to the standard
formalism. Berezin's method  can be seen as a particular case of deformation quantization  valid for
k\"ahlerian varieties, which permits to use coherent states to obtain the $*$--product \cite{PE86}.

In this paper  we pretend  to develop a quantum formalism for physical  systems with 
non-euclidean phase spaces.  We present the particular case of the cylinder as  phase space
with nontrivial topology.  Our approach is based on the well known formalism on
$\R^2$, in particular on the Weyl correspondence in $\R^2$ that it is carried to
the cylinder using the Weil--Brezin--Zak transform~\cite{WE64,BR70,ZA67}.

Since  to quantize means to establish a bijective correspondence between classical and
quantum observables, in principle it will be possible in any case to invert it and to
obtain a new correspondence assigning a function (symbol) on the phase space to each
observable or quantum state.  In this case the correspondence will be used to construct a
quantum formalism in phase space without  making use of the canonical quantization.
So, quantization and phase space formalism revel as two aspects of a same problem and
have to be studied simultaneously.

In the literature there exist several quantum formalisms on phase spaces related with the
presented here.  For instance, Berezin studied the cylinder quantization in
~\cite{BE74} showing that the ``admisible'' manifold  is a set of paralell circles, 
such that the distance
between two consecutive circles is $\hbar/2$.
Berry~\cite{BY77}, in a  work about the quantum rotor, and 
Mukunda~\cite{MK79}, in a study about angle variables, developed independently the same
formalism. 
Since their results gave a very complicated expression for the $*$-product,
Kasperkovitz {\it et al.\/}~\cite{KF87} introduced a modified formalism which describes
the motion of two particles on the circle. It was used by  Arnold {\it et
al.\/}~\cite{AD89} to  study the dynamics of  Bloch electrons.  Later,
Kasperkovitz and Peev~\cite{KP94} completed all these  ideas introducing some quantum
formalisms on phase spaces for systems with some kind of periodicity. In particular, they
considered systems where the position variable moves on a circle, which is the case
studied here. In ref.~\cite{AMO96,AO97} Arratia {\it et al.\/} used the Moyal quantization
method and profited the fact that the cylinder is a coadjoint orbit, to construct Moyal
quantizers.  More recently,  
 Pleba\'{n}ski and collaborators~\cite{PPTT00} constructed a quantization algorithm 
based on a family of unitary operators.

Thus, we can conclude that it seems necessary a more suitable framework to relate
the quantum formalisms on the plane and on the cylinder. We show in this work that
such a relation is easily realized by the so-called Weil--Brezin--Zak (WBZ) transform 
\cite{FO89} which, under different forms, has been used in the past for the study of
periodic potentials and other related problems \cite{RS78}. We will see that the WBZ
transform not only allows us to reproduce in a natural way all the known results for the
quantum formalism on the cylinder, but also provides us with a solution to  two important
topics: coherent states and quantization. We will also  prove in this work that
Kasperkovitz and Peev's formalism as well as Pleba\'{n}ski and collaborators formalism are
equivalent to ours.

The paper is organized as follows. In section~\ref{moyalquantization0} we
review  the foundations of the  quantizations via Weyl and Moyal approaches. 
In section~\ref{moyalquantization2} we quantize the cylinder as phase space using  
 the Weil--Brezin--Zak transform that allows us to translate the well known Weyl formalism
on $\R^2$.  Section~\ref{moyaljaremaquantizer}
contains a brief resume of the  quantum formalism introduced in
ref.~\cite{PPTT00}.  Some conclusions and remarks close the paper. In particular, 
we prove the equivalence of
our  Moyal quantization procedure with the Kasperkovitz--Peev~\cite{KP94} and  
Pleba\'{n}ski and collaborators~\cite{PPTT00} formalisms.

\sect{Moyal quantization\label{moyalquantization0}}

\subsect{Weyl mappings\label{weylquantization}}

One of fundamental obstacles which appear in a quantization procedure is establishing
some one-to-one relation between classical and quantum observables. When the classical
theory is formulated in terms of symplectic geometry and the quantum world is described in
a Hilbert space, to quantize means to find some bijection between a set of functions
and a collection of linear operators. 

It is usually assumed that in the case of $\R^2$
(generalization for $\R^{2n}$ is straight\-for\-ward) operators $\widehat{Q}$ and
$\widehat{P}$ representing canonical observables $q$ and $p$ respectively, are chosen like
in the canonical quantization programme, i.e. they are self-adjoint and  fulfill the
commutation relation
$$
[\widehat{Q},\widehat{P}]= i \hbar \widehat{1}.
$$
 Operators $\widehat{Q}$ and
$\widehat{P}$ are unbounded, which originates problems about definition domains and
self-adjoint extensions.   These troubles disappear if  unitary operators,  obtained by
exponentiation of $\widehat{Q}$ and $\widehat{P},$ are used.  
Weyl \cite{WY31} proposed to substitute the canonical quantization rules for the postulate
that the exponential function 
$e^{i(yq + xp)},\ x,y \in
\R$, has to be associated with the operator 
$e^{i(y\widehat{Q} + x\widehat{P})}$. In other words, whereas  the canonical quantization
determines some representation of the Heisenberg algebra by self-adjoint operators, the Weyl
quantization  gives a unitary   representation of the Heisenberg group, equivalent up
to a factor, to the  Schr\"odinger representation.

It is well known that any   tempered distribution $f$ can be written in terms of 
 exponentials by means of its  Fourier transform $\widetilde f$ on
$\R^2$, namely
\be\label{transformadainversa}
f(q,p) = \frac1{2\pi\hbar} \int_{-\infty}^\infty dy
\int_{-\infty}^\infty dx\, \widetilde f(y,x)\, e^{i(yq + xp)/\hbar} . 
\ee
The Fourier transform is defined by
\be
\widetilde f(y,x) = \int_{-\infty}^\infty dq
\int_{-\infty}^\infty dp\  f(q,p) e^{-i(yq + xp)/\hbar}.
\ee
This expression suggests the following  definition of the Weyl application 
\be\label{weylquantizations}
\widehat f = \frac1{2\pi\hbar} \int_{-\infty}^\infty dy
\int_{-\infty}^\infty dx\, \widetilde f(y,x)\, e^{i(y \widehat{Q}+ x\widehat{P})/\hbar}.  
\ee
After some computations one gets that in coordinate representation
\be
\label{pop1}
(\widehat f \p)(x) = \frac1{2\pi\hbar} \int_{-\infty}^\infty dy
\int_{-\infty}^\infty dp\, f\bigl(\frac12(x+y), p\bigr)\,
e^{i(x-y)p/\hbar}\, \p(y),     
\ee
with $\p \in {\cal S}(\R)$ (${\cal S}(\R)$ being the  Schwartz space). So, $\widehat f$
is an  integral operator with kernel 
$$
K_f(x,y) = \frac1{2\pi\hbar} \int_{-\infty}^\infty dp\,
f\bigl(\frac12(x+y), p\bigr)\, e^{i(x-y)p/\hbar},  
$$
which acts as $L({\cal H}): {\cal S}(\R) \rightarrow {\cal S}'(\R) $ (${\cal S}'(\R)$
denotes the space of tempered distribution)
\cite{FO89,VO77,VO78}.

From the last expression  one can obtain that
\be\label{1.53}
f(q,p) = \int_{-\infty}^\infty dy\, e^{iyp/\hbar}\,
K_f\bigl(q - \frac y2, q + \frac y2\bigr).  
\ee 
So, the tempered distribution $f(q,p)$ is determined by the kernel.

Relation (\ref{1.53}) is known as the Weyl correspondence.
Formulas (\ref{weylquantizations}) and (\ref{1.53}) are connected with  Weyl
ordering of operators. Generalization of them for other orderings  can be found in
\cite{Prz}.

\subsect{Moyal quantizer\label{moyalquantization1}}

Weyl application rule  
(\ref{weylquantizations}) does not have a suitable form 
to extend it for systems with phase spaces different from $\R^{2n},$
 since it is associated with the concept of  Fourier
transform. 

Let us come back to (\ref{pop1}) and introduce a new variable
  $q = \frac12(x+y).$ Now 
$$
(\widehat f \p)(x) = \frac1{\pi\hbar} \int_{-\infty}^\infty dq
\int_{-\infty}^\infty dp\, f(q,p)\, e^{i2p(x-q)/\hbar}\, \p(2q-x). 
$$
Defining the family of operators parametrized by $q,p \in \mathbb R$
\be\label{kernelsr2}
\bigl(\widehat \Om(q,p)\, \p\bigr)(x) := \frac1{\pi\hbar} e^{i2p(x-q)/\hbar}\,
\p(2q-x),   
\ee
we can rewrite the Weyl application form as
\be\label{operatorfuntion}
\widehat f = \int_{R^2} f(q,p)\, \widehat \Om(q,p)\, dqdp,  
\ee
where
\be
\label{pop2}
\widehat{\Om}(q,p) := \frac1{2\pi\hbar} \int_{-\infty}^\infty dy
\int_{-\infty}^\infty dx\, e^{-i(yq + xp)/\hbar}\, e^{i(y\widehat{Q} +
x\widehat{P})/\hbar}.
\ee
It is easy to see that in (\ref{operatorfuntion}) the term containing Fourier
transformations has been hidden in the symbol $\widehat{\Om}(q,p).$ It suggests that in
case of systems different from  $\R^{2n}$ to quantize them it is sufficient to define some
family of operators $\widehat{\Om}(q,p)$ and to substitute the integral over $\R^{2n}$ by
integration over the symplectic space $\Gamma$ of the system.

Operators $\widehat{\Om}(q,p)$ were introduced independently by  Grossmann \cite{GR76} and
Royer \cite{RO77}. In the past they were called  ``Stratonovich--Weyl
quantizers''.

Let us analyse some basic properties of $\widehat{\Om}(q,p)$ in $\R^{2}.$

Firstly, note that (see (\ref{kernelsr2}))
$$
\bigl(\widehat \Om(0,0)\, \p\bigr)(x) = \frac1{\pi\hbar}\, \p(-x)   
$$
represents (up to a factor) a reflection with respect to the origin. Hence, the operators
$\widehat{\Om}(q,p)$ can be seen as parity operators on phase spaces. 
Moreover, it is possible to prove that 
they are
self-adjoint and the map $\R^2   \to \widehat \Om(q,p)$ is
bijective. 

It has been showed  \cite{ST56} that
\be\begin{array}{lrl} \label{propertiesSW}
&\tr \widehat \Om(q,p) =& 1, \qquad \forall q,p \in \R,    \\[0.30cm]
&\tr\{\widehat \Om(q,p)\, \widehat \Om(q',p')\} = &\delta(q-q')\, \delta(p-p'),
\end{array}\ee 
where both of expressions have to be seen in the sense of the distributions.
The first expression of  (\ref{propertiesSW}) is equivalent to the statement  that the constant
function equal to 1 is associated with the identity operator.
 The second
property, called  traciality,
 allows us to rewrite the formula for the
 Weyl correspondence as
\be
\label{pop3}
f(q,p) = \tr\{\widehat f\, \widehat \Om(q,p)\} . 
\ee
 Moreover, after some
computations one obtains
\be\label{tracefg}
\tr\{\widehat f\, \widehat g\} = \int_{\R^2} f(q,p)\, g(q,p)\, dqdp.  
\ee

The Moyal product of two functions  on phase space is defined by 
$\widehat{(f\star g) } := \widehat f\, \widehat g.$ It can be computed from 
(\ref{kernelsr2}) and (\ref{pop3}) by the
following formula
$$
(f \star g)(q,p) = \int_{\R^2} dq'dp' \int_{\R^2} dq''dp''\, 
\tr\{\widehat \Om(q,p)\, \widehat \Om(q',p')\, \widehat \Om(q'',p'')\}\,
f(q',p')\, g(q'',p'').      
$$
The equality holds
$$
\int_{\R^2} (f\star g)(q,p)\, dqdp = \int_{\R^2} f(q,p)\,g(q,p)\, dqdp.  
$$

By  analogy with the case $\R^2$   some axioms have been established to be fulfilled by
operators $\widehat{\Om}(q,p)$ in a general situation.

Let  $\Gamma$ be an arbitrary phase space of a physical system, and  
$\h$ the  corresponding Hilbert space of  quantum states of this system.

The Moyal quantization is a map  $\widehat{\Om}: f \mapsto \widehat f$ among  functions (or
distributions) on
$\Gamma$ and operators  $L(\h)$, given by
\be\label{quantizando}
\widehat f = \int_{\Gamma} f(u)\, \widehat \Om(u)\, du,  
\ee
where $du$ is the volume determined by the symplectic form on $\Gamma.$ The operator
$\widehat \Om$ is the Moyal quantizer defined as a map from 
 $\Gamma$ in the set of operators  $L(\h)$, such that:
\begin{enumerate}
\item  $\widehat \Om$ is bijective,

\item $\widehat \Om(u)$ is self-adjoint for $\forall u \in \Gamma$,

\item $\tr\widehat \Om(u) = 1$  in the distributional sense for $\forall u \in \Gamma$,

\item Traciality:
\be\label{tracialidad}
\int_{\Gamma} \tr\{\widehat \Om(u)\, \widehat \Om(v)\}\, 
\widehat \Om(v)\, dv = \widehat \Om(u)     
\ee
for $\forall u \in \Gamma$.
\end{enumerate}
For elementary physical systems with a symmetry group $G$ their phase spaces $\Gamma$ shall
be, in general,  coadjoint orbits (homogeneous spaces) of  $G$ and $\h$ will carry a
unitary irreducible representation
$U$ of $G$ associated with $\Gamma$.    The kernel $\widehat\Om$ should verify an
additional condition (covariance property)
$$
\widehat \Om(g \cdot u) = U(g)\, \widehat \Om(u)\, U(g^{-1}), \qquad
\forall u \in \Gamma, \ \forall g \in G.    
$$

The inverse mapping $\widehat{f} \rightarrow f$ is defined as follows
\be\label{formula1}
f(u) = \tr\{\widehat f\, \widehat \Om(u)\} .    
\ee
Besides  $\widehat 1 = I_{\h}$ and the traciality condition are equivalent to
\be\label{formula2}
\tr\{\widehat f\, \widehat g\} = \int_{\Gamma} f(u)\, g(u)\, 
du =\int_{\Gamma} (f\star g)(u)\, du,   
\ee
with  the Moyal product  defined by
\be\label{formula3}
(f \star g)(u) = \int_{\Gamma} dv \int_{\Gamma} dw\, 
\tr\{\widehat \Om(u)\, \widehat \Om(v)\, \widehat \Om(w)\}\, f(v)\, g(w).
\ee

Note that the weak point of this quantization method is the lack of
general theorems about  existence and uniqueness of the Moyal quantizer.

\sect{Moyal quantization on the cylinder via Weil--Brezin--Zak
transform\label{moyalquantization2}}

\subsect{The Weil--Brezin--Zak (WBZ) transform\label{WBZ-transform}}

The Weil--Brezin--Zak (WBZ) transform  is well known in solid state 
theory where it gives rise to the Bloch functions \cite{RS78}. It   is a unitary map 
$T: L^2(\R) \to L^2(S^1\times S^{1*})$ defined by  
\be \label{WBZ}
(T\p)(q,k) = \sum_{n=-\infty}^\infty e^{inak}\, \p(q-na),   
\ee
where $S^{1*}=[0, 2\pi/a)$ is the dual space of $S^1 = [0,a)$, $\p \in L^2(\R),\ q
\in S^1$ and $k \in S^{1*}$. Functions $T\p(q,k)$ are periodic in $k$ and
quasiperiodic in $q$
\be (T\p)(q+na, k+m\frac{2\pi}a) = e^{inak}\, (T\p)(q,k), \qquad  n,m \in \Z.    
\ee
Fixing a value of $k$ the operator $T$ (hereafter
denoted by $\k T$) is a projection onto $L^2(S^1)$. Thus, $\k\p = \k T \p$
is the decomposition of $\p$ and we obtain the constant fiber direct integral
decomposition \cite{RS78},
$$
L^2(\R) \cong \int_{S^{1*}}^\oplus dk\, L^2(S^1).
$$
Let  $A$ be an operator  on $L^2(\R).$  If there exists a function 
$A^{(\cdot)}\colon S^{1*} \to {\cal L}(L^2(S^1))$ such that
$$
\k{(A\psi)} = \k A \k\psi, \qquad \forall \psi \in L^2(\R),
$$
it is said that $A$ is ``decomposable''and we can write
$$
A = \int_{S^{1*}}^\oplus dk\, \k A.
$$
Operators $\k A$ are called the ``fibers'' of $A$. 

It is easy to prove that an operator $A$ in $L^2(\R)$ is decomposable if and only if
it  is invariant  under the translation $T_a$, i.e., $T_a$ A $T_a^{-1}=A$.

The definition of decomposable operator works for bounded operators but it is possible
to extend it to unbounded self-adjoint operators \cite{RS78}.

If $\k A$ is
self-adjoint for each $k$, we have
$$
F(A) = \int_{S^{1*}}^\oplus dk\, F(\k A)     
$$
for any bounded function $F$ on $\R$.

Examples of decomposable operators are: the momentum operator $\widehat{P}$, the
translation operator $T_x={\rm exp}\; (i x \widehat{P}/h)$ and $E={\rm exp}\; (i2\pi
\widehat{Q}/a).$

\subsect{Weyl correspondence via WBK transform\label{weylquantizationcilindro}}

 The approach to  quantization on the cylinder 
 presented in this subsection  is based on the fact that $L^2(\R)$ is isomorphic to
$L^2(S^1\times S^{1*})$.

From this point of view, decomposition of operators can be considered as the natural link
between quantum theories on the circle and on the real line. That is because decomposable
operators are operators on $L^2(\R)$ which are invariant under the translation 
of length $a$,
so they are really acting on $L^2(S^1)$. This is the case of operators
$\widehat f$ associated with periodic functions in the position variable (we can of course
consider $f$ as a function (or distribution) on the phase space 
$S^1\times\R$) \cite{FO89}. 

The WBZ transform allows us to quantize on $S^1\times\R$ in
a  natural way as a particular case of quantization on $\R \times \R$.  
Functions $f(q,p)$ defined  on the cylinder are periodic in the position variable $q \in
[0,a).$   The Weyl quantization on
$S^1 \times \R$ is carried by  the fibers $\k{\widehat f}$ of the operator $\widehat f$,
 defined by (\ref{weylquantizations}).
 Since in this case
$f(q,p)$ is periodic in $q,$   operator  ${\widehat f}$ yields
\be
\widehat f = \frac1{2\pi\hbar} \int_{-\infty}^\infty dx
\sum_{n=-\infty}^\infty \widetilde f_n(x)\, e^{i(2\pi n\widehat{Q}/a +
x\widehat{P}/\hbar)},
\ee
where
\[
\widetilde f_n(x):= \int_{0}^{a} dq f(q,p) \exp{(-iqn/\hbar)}.
\]
This operator can be seen as an integral operator 
$$
(\widehat f \p)(x) = \int_{-\infty}^\infty K_f(x,y)\, \p(y)\, dy, \qquad
\p \in {\cal S}(\R),
$$
with integral kernel
\be\begin{array}{ll}\label{kernel}
\displaystyle 
K_f(x,y) &=   \displaystyle
\frac1{2\pi\hbar} \sum_{n=-\infty}^\infty
\widetilde f_n(y-x)\, e^{i\pi n(x+y)/a}     \\[0.30cm]
&=  \displaystyle 
\frac1{2\pi\hbar} \int_{-\infty}^\infty
f\bigl( \frac12 (x+y), p \bigr)\, e^{i(x-y)p/\hbar}\, dp.   
\end{array}
\ee
 From the second equation 
(questions about domains and convergence can be solved appealing to suitable conditions on
$f$ \cite{FO89}) we arrive to the following intermediate expression 
\be\begin{array}{ll}\label{fibra}
\k{(\widehat f \p)}(q) =&  \displaystyle
\frac1{2\pi\hbar} \int_{-\infty}^\infty dp \int_{-\infty}^\infty dx\,
f\bigl( \frac12(q+x), p \bigr)\\[0.40cm] 
&\qquad  \displaystyle \times e^{i(q-x)p/\hbar}
e^{iak\,\PE{x/a}}\, \k\p\bigl( (x)_a \bigr),   
\end{array}
\ee
where $x = (x)_a + a\PE{x/a}$ and $[y]$ the integer part of the real number $y$. 
Finally, we obtain the operator  in $S^1\times\R$  given by
\be \label{Weylcircle}
\k{\widehat f} = \frac1{2\pi\hbar} \int_{-\infty}^\infty dx \sum_{n=-\infty}^\infty
\widetilde f_n(x) e^{i\pi nx/a}\, E^n \exp \left(ix \k P /\hbar\right),
\ee
which can  be seen as an integral operator on
$L^2(S^1)$ (under suitable conditions on the function $f$ on $S^1\times\R$)
\be\label{intoperator}
(\k{\widehat f} \f)(q) =  \int_0^a \k K_f(q,q')\, \f(q')\, dq',
\ee
with  integral kernel
\be \label{intkernel}
\begin{array}{l}
\k K_f(q,q') =  \displaystyle
\bigl(\k T K_f(\cdot,q') \bigr)(q)\\[0.30cm]
\qquad
 =  \displaystyle
\frac1{2\pi\hbar} \sum_{n=-\infty}^\infty e^{inak}
\int_{-\infty}^\infty dp\, f\bigl(\frac12(q+q'-na), p\bigr)\, e^{i(q-q'-na)p/\hbar}.
\end{array}\ee

The operator  (\ref{Weylcircle}) has following properties:
\begin{enumerate}
\item  $\widehat f^{(k)\dagger} = \k{\widehat{f^*}}$,

\item  if $f$ is a function only of $p$, then $\k{\widehat f} = f(\k P)$ as defined
by the spectral theorem,

\item  if $g$ is a function only of $q$, we know that $g(q) =
{\sum_{n=-\infty}^\infty} \widetilde g_n e^{i2\pi nq/a}$. Then $\k{\widehat g} =
\sum_{n=-\infty}^\infty \widetilde g_n E^n$.
\end{enumerate}

An interesting fact is that the operators in $L^2(S^1)$ appearing in (\ref{Weylcircle})
$$
\k U(t,x,n) :=
e^{i\pi(2t+nx)/a}\, E^n\, e^{ix \k P/\hbar}
$$ 
verify the following statements:
\begin{enumerate}
\item 
It determines a unitary irreducible   representation
of  $\{\, (t, x,
\frac{2\pi}a \hbar n) \ |\  t,x \in \R, \ n \in \Z \,\}$,   subgroup of the 
Heisenberg group.

\item
The operators $\k U(x,n) := \k U(0,x,n)$ carry a unitary irreducible realization
(i.e. up to a factor representation) of the  subgroup
$\{\, (x, \frac{2\pi}a \hbar n) \ |\  x \in S^1, \ n \in \Z \,\}$
of the abelian group $S^1 \times \R$.
\end{enumerate}    

The proof can be done by considering
the  Schr\"odinger representation of the Heisenberg group  given by the operators
$$
U_{\lambda}(t,y,p) := e^{i\lambda tI} \exp\bigl(\frac i\hbar (p\widehat{Q} -
y\widehat{P})\bigr) , 
$$
where  $\lambda$ is a real parameter labelling the representation, and that
the operators $U_\lambda(t,y,p)$ are only  decomposable
when $p = 2\pi\hbar n/a$ with $\ n \in \Z$. The  
operators $\k U(t,x,n)$ are obtained  decomposing the operators
 $U_{2\pi/a}(t,-x,2\pi\hbar n/a)$. 

\subsect{Coherent states on the circle and Weyl quantization\label{coherentstates}}

Coherent states (CS) in $L^2(S^1)$  can be constructed~\cite{stephan93,GO96} by
decomposition of the standard Weyl--Heisenberg (WH) CS in $L^2(\R)$ \cite{PE86,KS85} 
\be \label{HWCS}
\eta_{q,p}(x) = \exp\bigl(\frac i\hbar  p  (x - \frac {q}2)\bigr)\,
\eta_0(x - q\,), \qquad x, q, p \in \R, \ \eta_0 \in L^2(\R),
\ee
by  the WBZ transform (\ref{WBZ}) 
\be
\k\eta_{q,p}\equiv \ket{q,p;k}:=\k T \eta_{q,p},\quad (q,p) \in S^1\times\R.
\ee
They verify the  resolution of unity
$$
\frac1{2\pi\hbar} \int_0^a dq\int_{-\infty}^\infty dp\, \ket{q,p;k}
\langle q,p;k| =  \widehat{I}.
$$
Choosing the fiducial state $\eta_0 \in L^2(\R)$ to be a normalized
Gaussian, 
\be \label{fiducialgauss}
\eta_0(x) = \left(\frac\om{\pi\hbar}\right)^{1/4}
\exp\left(-\frac\om{2\hbar}x^2\right),
\ee
the CS's on the circle take the form 
\be\ba{l} \label{CS1}
\k\eta_{q,p}(q') = \1{1/4} 
\exp {\left(\frac i{2\om\hbar} pz^*\right)}\,
\exp {\left(-\frac1{2\om\hbar}(z^* - \om q')^2\right)}  \\[0.30cm]
\qquad\qquad\qquad\times \th\!\left(i\frac a{2\hbar}(z^*-\om q'-ik\hbar); 
\rho_1 \right)\!,     
\ea\ee
with $z^* = \om q + ip$, $\rho_1 =
\exp (-{a^2\om}/({2\hbar}))$ and $\th(z;\rho) = 
\sum\limits_{n=-\infty}^\infty \rho^{n^2} e^{2inz}$, $|\rho|<1$. By $\th(z;\rho)$ 
we denote the Theta
function (sometimes written as $\th_3$)~\cite{MU83, ER81}.

Since these CS have been constructed by decomposition of the standard
WH CS in
$L^2(\R)$, we can compare both kinds of CS. The physical properties of the
CS on the circle depend  on some dimensionless parameter, related to the spread of
the initial WH-CS. If it is smaller than the length $a$ of the circle, the CS
on the circle are very similar to the WH-CS. But when such spread is comparable or
bigger than $a$, the CS on the circle are rather like plane waves. A detailed study of 
 the properties of these CS  can be found in
\cite{GO98}. For instance, the expectation values of the position and 
momentum are obtained as well as the
Heisenberg uncertainty relation.  In
\cite{KR96} a particular case of these CS were introduced 
by Kowalski {\em et al} to study a quantum
particle on the circle.  Recently in
\cite{KR02} Kowalski  and  Rembieli\'nski have introduced new uncertainty relations 
for the position and
momentum on the circle using the above mentioned CS.

The Weyl
correspondence on the cylinder (\ref{Weylcircle}) 
also  presents a correct semiclassical behaviour \cite{stephan93, GO96} 
For this purpose  the CS on the
circle $\ket{q,p;k}$ are used for computing  the expectation values of
the operators in these states. Also  the limit of these expectation values  
when $\hbar$ goes  to $0$ is
taken. We display the final results (see \cite{GO96} for details). Most
of the computations have been  made using the functional equation of $\th$
\cite{MU83,ER81}. So, one previously gets
$$
\ba{l}
\lim_{\hbar\to 0} \langle q,p;k\ket{q,p;k} = 1,     \\[0.2cm] \displaystyle
\lim_{\hbar\to 0} \frac1{2\pi\hbar} |\langle q',p';k\ket{q,p;k}|^2 =
\delta(q'-q)\,\delta(p'-p) .
\ea  
$$
The second expression shows that the
CS $\ket{q,p;k}$ concentrates at the phase space point $(q,p)$ in the semiclassical
limit. The expectation values of the operators $E$ and $\k P$ are, respectively,
$$\ba{l}
\lim_{\hbar \to 0} \langle q,p;k|E\ket{q,p;k} = e^{i2\pi q/a},   \\[0.40cm]
\displaystyle
\lim_{\hbar \to 0} \langle q,p;k|\k P \ket{q,p;k} = p.   
\ea
$$
After some cumbersome but straightforward calculations, it is  proved
that, if $\k{\widehat f}$ is the operator associated by the Weyl application
(\ref{Weylcircle}) to the function $f(q,p)$ on $S^1\times\R$, then
$$
\ba{l}
\lim_{\hbar \to 0} \langle q,p;k|\k{\widehat f}\ket{q,p;k} = f(q,p),  \\[0.30cm]
\displaystyle
\lim_{\hbar \to 0} {1 \over i\hbar} \langle q,p;k|[\k{\widehat f},\k{\widehat g}]
\ket{q,p;k} = \{f,g\}(q,p),   
\ea
$$
where $\{f,g\}$ is the Poisson bracket of the functions $f,g$ on $S^1\times\R$.

It is worthy noting  that these CS have not been constructed by Perelomov's method
\cite{PE86}, as an orbit under a Lie group representation. They constitute a
non-trivial example of the ``reproducing triplets'' introduced in \cite{AA91}. In
reality, there is not a group generating these CS. This  is related to the fact that the
Schr\"odinger representation is not decomposable.

The action of the operators $\k U(x,n)$ over the 
CS $\ket{q,p;k}$ is
$$
\k U(x,n)\, \ket{q,p;k} = e^{i\pi nq/a} e^{ipx/(2\hbar)}\,
\ket{q-x, p + \frac{2\pi}a \hbar n; k}.      
$$

All that  has strong consequences on the physical interpretation of the
parameters $q$ and $p$. 
In the case of the  WH-CS the  parameters of the ``classical'' phase space
$\R^2$ are in correspondence with the expectation values of the operators $\widehat{Q}$ and
$\widehat{P}$.  The Schr\"odinger representation, generating these CS, acts by
translations over the expectation values, which  constitute a 
true ``quantum'' phase space  coinciding in this case with the  classical phase space.
However,  the CS on the cylinder, $\ket{q,p;k}$, are very different, since the
luck of a group generating these CS  does not allows us to give a clear physical 
interpretation  for $q$ and $p$. In fact, it is known that the cylinder cannot be seen
as a quantum phase space with translation operators like those of $\R ^2$ because the
discrete nature of the spectrum of the momentum operators in  $L^2(S^1)$. 
Only the  subgroup
$\{\, (x, 2\pi\hbar n/a) \ |\  x \in S^1, \ n \in \Z
\,\}$ of $S^1 \times \R$ can be  used as quantum phase space,
being $\k U(x,n)$ the corresponding translation operators. This result is in
agreement with the conclusions obtained by other authors, for instance~\cite{KP94,PPTT00}.  Obviously, the 
  parameters $q$ and $p$ do not coincide in general with the expectation values
of the position and momenta operators for the CS $\ket{q,p;k}$. 

\subsect{Weyl  symbols on the cylinder\label{weylsymbolscilindro}}

The  Weyl application (\ref{weylquantizations}) on $\R^2$ is one--to--one, so
it can be inverted. The function (or symbol) 
\[ 
f= Tr\{\widehat{f}\widehat{\Om}\}
\] 
is associated
with each operator $\widehat{f}$. These symbols are useful to develop the phase space formalism of
quantum theory \cite{GR74,AH81,gad}. However, the Weyl correspondence on the cylinder
(\ref{Weylcircle}) is not one--to--one, hence  symbols on
$S^1 \times \R$ cannot be defined in this way. 

Let
$\alpha_{n}$ be the  Fourier coefficients  of the function
$e^{-ik  q} \phi(q)$, where $\phi \in L^2(S^1)$ is an arbitrary function and  $k \in [0,2\pi /a)$ 
fixed,  
\be
\begin{array}{lll} \label{2.12}
\phi(q) &=&\displaystyle \sum_{n =-\infty}^{\infty} \alpha_{n}\,
e^{i(2\pi n/a +  k)  q} ,       \\[0.30cm]
\alpha_{n} &=& \displaystyle \frac1{a} \int_{S^1} d q\, \phi(q)\,
e^{-i(2\pi n/a +  k)   q} .       
\end{array}
\ee
There exists at least one function $\p \in
L^2(\R)$ such that $\phi = \k\p$, i.e.,  $\p$ will be some
function of $L^2(\R)$ verifying 
\be\label{2.13}
\alpha_{n} = \frac{1}{a}\,
\widetilde\p\bigl((\frac{2\pi}{a} n +  k)\hbar\bigr).    
\ee
On the other hand, we can write
\be
\begin{array}{l} \label{2.15} 
\displaystyle\frac{1}{a}\,
\widetilde\p\bigl(( 2\pi n/a+ k)\hbar\bigr) = \displaystyle\frac1{a}
\int_{R} d x\, \p( x)\, e^{-i (2\pi n/a +  k)   x}   \\[0.30cm]
\qquad \qquad =  \displaystyle\frac1{a} \sum_{ a \in \Z} \int_{S^1} d q\,
\p( q -  a)\, e^{i a   k}
e^{-i( 2\pi n/a +  k)   q}         \\[0.30cm]
\qquad\qquad =   \displaystyle\frac1{a}
\int_{S^1} d q\, \biggl(\sum_{ a\in\Z} e^{i a k}\,
\p( q -  a) \biggr) e^{-i( 2\pi n/a +  k)   q}       \\[0.30cm]
\qquad \qquad =  \displaystyle \frac1{a} \int_{S^1} d q\, \p( q)\,
e^{-i( 2\pi n/a +  k)   q} .     
\end{array}\ee
Hence, we have derived  the expression for the  Fourier coefficients of the function 
$e^{-i k  q}\k\p(q)$.
Since $\p$ verifies  relations  (\ref{2.13}), then
$\k\p = \phi$ due to the uniqueness of the Fourier coefficients.
There are many  functions   $\p$ that verify (\ref{2.13}) for some fixed
$\alpha_{2\pi n/a }$. In other words,
given  $\phi$, the function $\p$, such that $\k\p = \phi$, is not uniquely  
determined. The factor $e^{-i k  q}$ has been introduced because the  functions
$\k\p( q)$, obtained by using  the expression (\ref{WBZ}), are quasiperiodic in 
$ q \in \R$. Multiplying (\ref{WBZ}) by $e^{-i k  q}$
we obtain  functions strictly periodic. The usual Fourier analysis can be applied to
them.

The second formula of (\ref{2.12}) shows that two functions $\p_1, \p_2
\in L^2(\R)$ have the same image under $\k T$ if
$$
\widetilde\p_1\bigl((\frac{2\pi}a n + k)\hbar\bigr) =
\widetilde\p_2\bigl((\frac{2\pi}a n + k)\hbar\bigr),  \qquad
\forall n \in \Z.
$$
The integral kernel  $\k K_f(q,q')$ of a Weyl  operator  $\k{\widehat
f}$ in $L^2(S^1)$ is the image, as function of $q$, under $\k T$ of the integral
kernel $K_f$ of the corresponding Weyl operator $\widehat f$ in $L^2(\R)$.
Then, two functions $f$ and $g$ on $S^1\times\R$ will be associated with the same
operator on $L^2(S^1)$ if
$$
\widetilde K_f^{1)}\bigl((\frac{2\pi}a n + k)\hbar,\, q'\bigr) =
\widetilde K_g^{1)}\bigl((\frac{2\pi}a n + k)\hbar,\, q'\bigr),  \qquad
\forall n \in \Z,\ q' \in S^1, 
$$
where $\widetilde K_f^{1)}$ denotes the partial  Fourier
transform of $K_f$ with respect to its first  variable. After a cumbersome computation we get 
$$
\begin{array}{lll}
 \widetilde K_f^{1)}\bigl((\frac{2\pi}a n + k)\hbar,\, y\bigr) &=&
\displaystyle
\frac{1}{a}\, e^{i(2\pi n/a - k)y} \sum_{m=-\infty}^\infty
e^{-i2\pi my/a}  \\[0.30cm]
 & &\qquad  \displaystyle
\times \int_0^a dq\, f\bigl(q,\, (\frac{\pi}am +
k)\hbar\bigr)\, e^{i2\pi(m-2n)q/a}.    
\end{array}
$$
Hence,
$$
\k{\widehat f} = \k{\widehat g} \iff
f\bigl(q,\, (\frac{\pi}an+k)\hbar\bigr) =
g\bigl(q,\, (\frac{\pi}an+k)\hbar\bigr),  \quad
\forall n \in \Z,\ q \in S^1 . 
$$
Thus, it is sufficient that  $f(q,p)$ and $g(q,p)$ coincide on a discrete set
of values of $p$, depending on  $\hbar$ to be the symbols of an operator on the cylinder. This observation
is basic for a correct definition of the symbols associated with operators on the cylinder.

It is possible to
define symbols of the operators $\k{\widehat f}$ for a quantum phase
space (the cylinder) different from the classical one as it is showed in the
following~\cite{GO96}. 

The integral kernel
(\ref{intkernel})  of $\k{\widehat f}$ can be rewritten as
\be \label{series}
\k K_f(q,q') = e^{ik(q-q')} \sum_{n=-\infty}^\infty e^{i\pi(q-q')n/a}\,
F(q+q',\, n),
\ee
where
\be \label{symbol}
F(x,n) := \frac1{2a} \left\{ f\bigl(\frac x2, (\frac\pi a n+k)\hbar\bigr) +
(-1)^n\, f\bigl(\frac{x+a}2, (\frac\pi a n+k)\hbar\bigr) \right\}.
\ee
From relation (\ref{series}) we deduce  that  there is a one--to--one
correspondence between the operators $\k{\widehat f}$ with integral kernel  
$\k K_f(q,q')$ and the functions $F(x,n)$. 
Hence, $F(x,n)$ is taken as the Weyl symbol of $\k{\widehat f}$. To see  this
relationship more clearly we can rewrite (\ref{series}),  using the new variables 
$x := q + q'$, $2y := q - q'$, as
$$
\k K_f(\frac x2 + y,\frac x2 - y) = e^{i2ky} \sum_{n=-\infty}^\infty
e^{i2\pi yn/a}\, F(x,n).
$$
This equation can be inverted   
\be\label{5.6}
F(x,n) = \frac1a \int_0^a dy\, e^{-i2(\pi n/a + k)y}\,
\k K_f(\frac x2 + y,\frac x2 - y).       
\ee
Note
that the symbol $F(x,n)$ only depends  on the value of the function
$f(q,p)$ for a discrete set of values of $p$. The  (non bijective) correspondence between
$F(x,n)$ and
$f(q,p)$ 
 given by (\ref{symbol}) can be seen as an alternative expression of the Weyl quantization
rule on the cylinder. 

The  property
\be\label{propiedadf}
F(x + a,n) = (-1)^n\, F(x,n),
\ee
allows us to consider the set $\{\, (q,n) \ | \ q \in S^1, \ n \in \Z \,\}$ as
the quantum phase space where the symbols are defined.
It is worthy noticing that the symbols $F(x,n)$ are periodic  in $x$ with period $a$ when $n$
is odd and period $2a$ when $n$ is even, but the classical functions $f(q,n)$ are periodic
in $q$ with period $a$. 


\subsect{The Moyal quantizer\label{moyalquantizationWBZ}}

As we mention before the basic problem of  Moyal's formalism is the existence of 
the Moyal quantizer $\widehat\Om$. It is usually  solved using group theory, but this method
does not work in the case of the cylinder because we have not a symmetry group. However, the existence
of the Weyl correspondence allows us to construct  $\widehat\Om$ and reformulate the
Weyl quantization as  Moyal quantization~\cite{GO96}. 

According to the definition of the Moyal quantization
(expression  (\ref{quantizando})) in the case of the cylinder we  write
$$
\k{\widehat f} = \int_{-\infty}^\infty dp \int_0^a dq\, f(q,p)\,
\k{\widehat\Om}(q,p),
$$
with $\k{\widehat f}$ being the operator associated with the classical function 
$f(q,p)$ by the Weyl correspondence (\ref{1.53}) and $\k{\widehat\Om}$  
the cylinder Moyal quantizer.  But the Weyl correspondence does not verify the bijectivity
requirement, so we are compelled to use the symbols $F(q,n)$ instead the classical
functions $f(q,p)$.  Hence, we have
\be\label{fibracylinder}
\k{\widehat f} = \sum_{n=-\infty}^\infty \int_0^a dq\, F(q,n)\,
\k{\widehat\Om}(q,n).
\ee
So, $\k{\widehat\Om}$ has to be defined on the
quantum phase space $S^1\times \Z$ as it happens with the symbols $F$. 

Although the operators $\k{\widehat f}$ were constructed by decomposition of the Weyl
operators in $L^2(\R)$, the corresponding Moyal quantizer cannot be obtained in the same
way since the Grossmann--Royer operators (\ref{kernelsr2}),
that determine the Moyal quantizer in $\R^2$,  are not decomposable.

Starting from expression (\ref{fibra}) and after some calculations
we obtain
\be \label{swkernel}
\left( \k{\widehat\Om}(x,n)\, \f \right)(q) = e^{ik(2q-x)}\, e^{i\pi(2q-x)n/a}\,
\f_{c}(x-q),   \qquad   \f \in L^2(S^1),
\ee
where  
 $\f_{c}(x) = e^{iak\,\PE{x/a}}\, \f\bigl(x - a\PE{\frac xa}\bigr)$.

We must check if the
operators (\ref{swkernel}) fulfill the conditions to be a Moyal quantizer displayed in
section~(\ref{moyalquantization1}).  The demonstration of that $\k{\widehat\Om}$ is  bijective and
self-adjoint is easy and we let it for the reader. The  proof of that  $\tr\widehat \Om(u) = 1$ 
and traciality (\ref{tracialidad})
requires to use an
appropriate orthonormal basis of
$L^2(S^1)$, $\bigl\{\,
\ket{n;k} \ \bigm| \ n \in \Z, \  k \in (S^1)'\ {\rm fixed} \,\bigr\}$, given by
$$
\langle q  \ket{n;k} = \frac1{\sqrt a} \exp\bigl[i(\frac{2\pi}a n + k)q\bigr].
$$
After a straightforward
computation  we get
\be\label{traza01}
\tr \k{\widehat\Om}(x,n) = \frac12 \bigl(1 + (-1)^n\bigr),
\ee
and 
\be\label{traciality}
\tr\{\k{\widehat\Om}(x,n)\, \k{\widehat\Om}(y,m)\} =
a\, \delta_{n,m}\, \delta(x-y)     
\ee
Note that $\k{\widehat\Om}$ is not a proper Moyal quantizer since the property of trace
unit is not accomplished.

It is worthy to mention that since
 $\k{\widehat f}$ is an integral operator with kernel
  $\k K_f$ and supposing that it is of trace class, we can prove that 
\be\label{simbolof1}
\tr \k{\widehat f} = \int_0^a dq\, \k K_f(q,q) = \sum_{n=-\infty}^\infty
\int_0^a dq\, F(2q,n)   ,
\ee
 and 
\be\label{simbolof2}
\tr \k{\widehat f} = \sum_{n=-\infty}^\infty \int_0^a dq\, F(q,n)\, \tr
\k{\widehat\Om}(q,n)  = \sum_{n=-\infty}^\infty \int_0^a dq\, F(q,2n).
\ee

On the other hand, the symbol associated  with the function $f \equiv 1$, i.e.
to the unit  operator in $L^2(S^1)$, is
\be\label{simbolo1}
F_1(n) = \frac1{2a} \bigl(1 + (-1)^n\bigr) .    
\ee
This result is due to the fact that $\tr \k{\widehat\Om}\neq 1$. 
The classical function $f \equiv 1$ is associated with the identity operator although
 its corresponding  quantum symbol is not 1. This last result  is intrinsic of this
formalism.   

Despite $\tr\k{\widehat\Om}\neq 1$ a quantum formalism can be developed in the cylinder as
phase space. The more relevant formulas for it are
(\ref{formula1})--(\ref{formula3}) and only those that dependent on  
$\tr\k{\widehat\Om}$ have to be modified. Thus, formula  (\ref{formula1}) as well as the
first equality of (\ref{formula2}) are valid, since they are only based on the tracial
property (\ref{traciality}). We have in these cases
\be\label{simbolo}
F(q,n) = \frac1a \tr \{ \k{\widehat f}\, \k{\widehat\Om}(q,n) \} ,
\ee
\be\label{trazados}
\tr \{ \k{\widehat f}\, \k{\widehat g} \} = a \sum_{n=-\infty}^\infty
\int_0^a dq\, F(q,n)\, G(q,n),   
\ee
with $F$ and $G$ the symbols of $\k{\widehat f}$ and $\k{\widehat g}$,
respectively.  Likewise, the  expression of the twisted product of two symbols
(\ref{formula3}) is directly applicable in this
case. With the help of formula (\ref{simbolo}) we obtain
\be\begin{array}{lll}
 (F \star G)(x,n)
&=& \displaystyle
\frac1a \tr \{ \k{\widehat f}\, \k{\widehat g}\, \k{\widehat\Om}(x,n) \} \\[0.40cm]
&=& \displaystyle 
\frac1a \sum_{m=-\infty}^\infty \sum_{l=-\infty}^\infty \int_0^a dy
\int_0^a dz\,  F(y,m)\, G(z,l)  \\[0.30cm]
& & \displaystyle
\qquad \times \tr \{ \k{\widehat\Om}(x,n)\, \k{\widehat\Om}(y,m)\,
\k{\widehat\Om}(z,l) \}  ,        
\end{array}\ee
where
\be\begin{array}{l}\label{trikernel}
\tr \{ \k{\widehat\Om}(x,n)\,\k{\widehat\Om}(y,m)\, \k{\widehat\Om}(z,l) \} \\[0.30cm]
\qquad  =  \frac12 \bigl(1 + (-1)^{n-m+l}\bigr) e^{i\pi(m-l)x/a}\, e^{i\pi(l-n)y/a}\,
e^{i\pi(n-m)z/a}.   
\end{array}\ee
However, the second equality of (\ref{formula2}) is not valid and it has to be
substituted by 
\be\begin{array}{lll}\label{alternativa1}
 \tr \{ \k{\widehat f}\, \k{\widehat g} \} &=& 
\displaystyle \sum_{n=-\infty}^\infty
\int_0^a dq\, (F \star G)(q,n)\, \tr \k{\widehat\Om}(q,n)   \\[0.35cm]
&= & \displaystyle
\sum_{n=-\infty}^\infty \int_0^a dq\, (F \star G)(q,2n), 
\end{array}\ee
where we have used (\ref{traza01}). Expression  (\ref{alternativa1}) as well as
(\ref{simbolof2}) are related to the fact that the symbol
$F_1(n)$, given by (\ref{simbolo1}), plays the role of unit function in the quantum
formalism on phase space. Note that  $\tr \k{\widehat\Om}(q,n) =a F_1(n)$.

\sect{Moyal quantization on the cylinder via unitary operators
\label{moyaljaremaquantizer}}

The quantization procedure, that we present  now, has been developed in
ref.~\cite{PPTT00}. It is based on results of
\cite{BY77,MK79} and we will call it the PPTT formalism.  Let
$(\theta,p)$ be the coordinates on the cylinder:
$\theta
\in[-\pi,\pi)$ and $p \in (-\infty,\infty)$ (the angular momentum 
 conjugate to  $\theta $).

Let us consider  the family of unitary
operators 
\be\label{baseunitaria}
\label{1.2j}
\widehat{U}(\tau,m)=\exp \left\{
i \left(
\frac{\tau}{\hbar}\widehat{p} + m \widehat{\theta}
\right)
\right\}, \;\;\; m \in \Z, \; -\pi \leq \tau < \pi.
\ee
The following property holds
\be
\label{2j}
Tr \left\{
\widehat{U}^+ (\tau,m) \widehat{U}(\tau',m')
\right\}= 2 \pi \delta_{m,m'}\delta^{(S)}(\tau -\tau'),
\ee
where $\delta^{(S)}(\theta)$ denotes the Dirac delta on the circle $S^1$.

Let $f=f(\theta,p)$ be a function on the cylinder and $\widehat{f}$ some linear 
operator representing  $f$ in the quantum case.  Expanding $\widehat{f}$ with
respect to the basis (\ref{baseunitaria})   one
gets
\be
\label{3j}
\widehat{f}= \frac{1}{(2\pi)^2} 
\sum_{m= - \infty}^{\infty}\int_{-\pi}^{\pi}\tilde{f}(\tau,m) \widehat{U}(\tau,m) d \tau,
\ee
where $\tilde{f}(\tau,m)$ is a function to be determined.
From (\ref{3j}) and (\ref{2j}) one obtains that
\be
\label{4j}
\tilde{f}(\tau,m)= 2 \pi Tr \left\{
\widehat{U}^+ (\tau,m) \widehat{f}
\right\}.
\ee
A consequence of expression (\ref{4j}) is that the
 operator $\widehat{f}$  determines a function $\tilde{f}(\tau,m)$ only for  
$\tau \in [-\pi,\pi)$.
It means that
if it is assumed (by analogy with $\R^2$) that $\tilde{f}(\tau,m)$ is 
the Fourier transform of $f(\theta,p),$ one is not  able to extract all the
information about this transform from (\ref{3j}) but only its values in the interval 
$[- \pi , \pi)$.

Considering the following distribution on the cylinder $  S^1 \times \R$
\be
\label{5j}
F(\theta,p):=
\sum_{n=-\infty}^{\infty}f(\theta,n \hbar)\delta(p-n\hbar),
\ee
its Fourier transform, $\widetilde{F}(p,\theta)$, is defined by
\be\begin{array}{lll}\label{6j}
\widetilde{F}(\theta,p)(\tau,m) &=& \displaystyle
\int_{-\infty}^{\infty}dp 
\int_{- \pi}^{\pi} d \theta F(p,\theta) 
\exp \left\{-i \left(\frac{\tau}{\hbar}p + m \theta\right)\right\}\\[0.30cm]
&=&  \displaystyle
\sum_{n=-\infty}^{\infty} \int_{- \pi}^{\pi} d \theta f(\theta,n \hbar)
\exp\{-i(\tau n + m \theta)
\}.
\end{array}\ee

The quantization rule (\ref{3j}) applied 
together with (\ref{5j}) and (\ref{6j}) constitutes a  one-to-one correspondence between
functions on $  S^1 \times \hbar \Z \subset   S^1 \times \R$ 
and operators on the Hilbert space $L^2(S^1).$

Substituting the second equality of (\ref{6j}) in (\ref{3j}) it is obtained that
\be
\label{7j}
\widehat{f}= \sum_{n=-\infty}^{\infty} \int_{- \pi}^{\pi} 
\frac{d \theta}{2 \pi} f(\theta,n \hbar) \widehat{\Omega}( \theta,n),
\ee
where
\be
\label{8j}
\widehat{\Omega}(\theta,n) :=  \frac{1}{2 \pi} 
\sum_{m=-\infty}^{\infty} \int_{- \pi}^{\pi} d \tau 
\exp\{-i(\tau n + m \theta) \widehat{U}( \tau, m),
\ee   
or, in an alternative form
\be
\label{8.1j}
\widehat{\Omega}(\theta,n)=\int_{- \pi}^{\pi} \exp(i \tau n) | 
\left[\theta + \frac{\tau}{2} \right] \rangle  \langle
\left[ \theta - \frac{\tau}{2}\right] | d \tau.
\ee
The operator $\widehat{\Omega}(n, \theta)$ is a Moyal quantizer since it verifies all the
required conditions. Notice that it is defined not on the whole cylinder $S^1 \times \R$
but only on
$S^1 \times  \Z.$

Obviously, from (\ref{7j}) one recovers $f$
\be
\label{10j}
f( \theta, n \hbar)=Tr \left\{\widehat{\Omega} \widehat{f} \right\}. 
\ee

The $*$--product of functions on the cylinder associated with this quantizer is given by
\[\begin{array}{l}\label{11j}
(f_1 * f_2)( \theta,n \hbar)= 
Tr \left\{\widehat{\Omega}( \theta,n) \widehat{f}_1   \widehat{f}_2\right\}
=  \displaystyle
\frac{1}{4 \pi^2} \sum_{n',n''=-\infty}^{\infty} \int_{- \pi}^{\pi} d \theta'
\int_{- \pi}^{\pi} d \theta'' \\[0.30cm]
\quad \times 
\left\{
f_1( \theta', n' \hbar)
Tr \left\{\widehat{\Omega}( \theta, n)\widehat{\Omega}( \theta',n')\widehat{\Omega}(
\theta'',n'')
\right\}
f_1( \theta'',n'' \hbar)\right\},
\end{array}\]
where $f_1( \theta, n \hbar)$ and $f_2( \theta, n \hbar)$ are arbitrary functions on 
$S^1 \times  \Z $ and  $\widehat{f}_1,\widehat{f}_2 $ their corresponding operators.
The explicit expression of the trikernel is
$$\begin{array}{l}\label{12j}
 Tr \left\{\widehat{\Omega}( \theta, n)\widehat{\Omega}( 
\theta',n')\widehat{\Omega}( \theta'', n'') \right\} =\exp\{2i[(n''-n)(\theta' - \theta)-
(n'-n)(\theta'' - \theta)] \}\nonumber
\\[0.30cm]
\qquad \times \{ 1 + sgn(\cos(\theta'' - \theta))sgn(\cos(\theta' - \theta))  
 + sgn(\cos(\theta' - \theta))sgn(\cos(\theta'' - \theta')) \nonumber \\[0.30cm]
\qquad\qquad + sgn(\cos(\theta'' - \theta'))sgn(\cos(\theta'' - \theta))\} ,
\end{array}$$
where
$
sgn(x)$ is equal to $1,0$ or $-1$ if $ x>0,  x=0$ or $x <0$  respectively.
Then, it is obtained the following expression for $f_1 * f_2$  
\be
\label{13j}
(f_1 * f_2)( \theta, n \hbar)=
\left\{
f_1( \theta, p) \exp\left(
\frac{i \hbar}{2} \stackrel{\leftrightarrow}{\cal P}
\right)f_2( \theta,p)
\right\}\big|_{p=n \hbar},
\ee
with\be
\label{14j}
 \stackrel{\leftrightarrow}{\cal P} \;=
\frac{\stackrel{\leftarrow}{\partial}}{\partial \theta }
\frac{\stackrel{\rightarrow}{\partial}}{\partial p}-
\frac{\stackrel{\leftarrow}{\partial}}{\partial p } 
\frac{\stackrel{\rightarrow}{\partial}}{\partial \theta}.
\ee
The same result was obtained by Berezin~\cite{BE74}.

\sect{Conclusions and remarks}

As we have mentioned before there exist other quantization procedures to be applied to the cylinder as
phase space. We shall prove that the methods proposed in ref.~\cite{KP94},\cite{PPTT00} and our attempt are
equivalent.

\subsect{Moyal quantization by WBZ  versus PPTT formalism}

The crucial role in the quantization procedure proposed in \cite{PPTT00} is played by a Moyal
quantizer.  Hence, to prove the equivalence of the quantum formalism introduced in ref.~\cite{PPTT00} and
ours it is enough to demonstrate   that formulas (\ref{fibracylinder}) and
(\ref{7j})   are equivalent.

The action of the Moyal quantizer (\ref{8.1j}) on an arbitrary function $\varphi(\alpha)$
is
\be
\label{ja1}
\left( \widehat{\Omega}( \theta,n) \varphi \right) (\alpha) = \exp\left( 
2 i n (\alpha - \theta) \right) \varphi(2 \theta - \alpha).
\ee
Substituting $ 2 \theta$ by $\chi$ (now $\chi \in [-2 \pi,2 \pi )$) we see that
(\ref{ja1}) equals
\be 
\label{ja2}
\exp\left(in(2 \alpha- \chi ) \right) \varphi(\chi - \alpha) .
\ee
This result (\ref{ja2}) can be also obtained by the quantizer $\widehat{\Omega}^{(0)}(n, x)$ given
by (\ref{swkernel}) with $k=0$ and  $a=2\pi$.

Now from expressions (\ref{7j}) and (\ref{ja2}) we obtain
\be
\label{ja3}
\widehat{f}= \sum_{n = -\infty}^{\infty} \int_{-2 \pi}^{2 \pi} \frac{d x}{4 \pi}
\widehat{\Omega}^{(0)}( x,n) f(\frac{x}{2}, n \hbar).
\ee
Splitting the above integral in a sum of three terms 
\[\begin{array}{lll}\label{ja4}
\widehat{f} &=& \displaystyle
\sum_{n = -\infty}^{\infty} \int_{- \pi}^{ \pi} \frac{d x}{4 \pi}
\widehat{\Omega}^{(0)}( x,n) f(\frac{x}{2}, n \hbar) \\[0.30cm]
& &\quad \displaystyle
+ \sum_{n = -\infty}^{\infty} \int_{-2 \pi}^{- \pi} \frac{d x}{4 \pi}
\widehat{\Omega}^{(0)}( x,n) f(\frac{x}{2}, n \hbar) \\[0.30cm] 
& & \quad  \displaystyle
+ \sum_{n = -\infty}^{\infty} \int_{ \pi}^{2 \pi} \frac{d x}{4 \pi}
\widehat{\Omega}^{(0)}( x,n) f(\frac{x}{2}, n \hbar),
\end{array}\]
using the property (see expression ({\ref{swkernel}))
\be
\widehat{\Omega}^{(0)}( x \pm 2 \pi,n ) = (-1)^n \widehat{\Omega}^{(0)}( x,n),
\ee
and the fact that the period of the function $f(x,n \hbar)$ is $2 \pi$, we finally arrive
to
\be
\label{ja5}
\widehat{f}= \sum_{n = -\infty}^{\infty} \int_{- \pi}^{ \pi} \frac{d x}{4 \pi} 
\widehat{\Omega}^{(0)}( x,n) \left( f(\frac{x}{2},n \hbar) + (-1)^n f(\frac{x}{2} +\pi ,n
\hbar) \right).
\ee 
This expression (\ref{ja5}) is equivalent to (\ref{fibracylinder}) when $k=0$, $a=2\pi$ and
$F(x,n)$ is given by (\ref{symbol}).

\subsect{Moyal quantization by WBZ  versus Kasperkovitz--Peev formalism}

The quantum formalism introduced by  Kasperkovitz and Peev  \cite{KP94} is  based on the Weyl
quantization.   

Let us start from formula (\ref{5.6}). Taking  $k = 0$ (i.e. periodic boundary conditions),
$a = 2\pi$ and making the variable change $y' = y + x/2$ we obtain 
formula (110) of \cite{KP94}
$$
A^{II}(Q,Y) = \int_0^{2\pi} dy'\, e^{iQ(Y-2y')}\, K_f^{(0)}(y', Y-y'), 
$$
with the identifications  
\[ 
A^{II}(Q,Y) = 2\pi\, F(Y,n) ,  \qquad Q = \pi n/a = n/2 .
\] 
Thus, our  Weyl symbol (\ref{symbol}) coincides with the function  $A^{II}$, 
introduced in
\cite{KP94}. Also  $A^{II}$ verifies a  condition analogous to  (\ref{propiedadf}). It was obtained
after   cumbersome  computations, and it is necessary to put by hand a period   of $4\pi$. However,
in our case expressions (\ref{5.6}) and (\ref{propiedadf}) have been
easily obtained from the properties of a quantization method   well defined for the
cylinder like is the  correspondence   (\ref{Weylcircle}). Our formalism 
and the formalism developed by Kasperkovitz and Peev are both based on the
  Weyl correspondence on $\R^2$. In particular, the   formal similarity that exists
between  expressions (\ref{1.53}) and (\ref{5.6}) (when $k=0$,
$a=2\pi$) explains the fact that it is possible a construction like has been made in
\cite{KP94}. Moreover, these expressions show that such a comparison can be only made
after introducing the double period $4\pi$.

It is worthy noting that the Weyl correspondence on the cylinder directly
distinguishes between the `classical' phase space $S^1\times \R$ and the `quantum'
phase space $S^1\times \Z$. The second one is the only one that admits
a Moyal formalism based on the use of symbols. The correct way of
defining these symbols comes from the properties of the Weyl
correspondence on the cylinder.

Some features of this formalism that  had not been sufficiently
explained in the literature appear in our case as simple consequences
of the method.

The Moyal quantizer that we have obtained has trace different from one.
This fact is related with the  use of symbols of functions
instead of function themselves. However, with some small modifications
our formalism works well and  is equivalent to other procedures
as we have showed above.


\section*{Acknowledgements}  
\noindent 
We acknowledge Stephan de Bi\`evre for his fruitful commentaries. This work has been
partially supported by DGES of the  Ministerio de Educaci\'on y Cultura of Spain under
Project PB98-0360, by the DGI of the Ministerio de Ciencia y 
Tecnolog\'{\i}a of Spain (project BMF2002-02000) and the Programme FEDER of the European
Community,  and the Junta de Castilla y Le\'on (Spain). J. T. acknowledges
la Secretar\'{\i}a de Estado de Educaci\'on y Universidades del Ministerio de Educaci\'on
y Cultura of Spain for the grant (SB2000-0129)  that supports his stay in the Universidad
de Valladolid.

\bigskip



\end{document}